\documentclass[prd,nofootinbib,showpacs,preprint]{revtex4}

\usepackage[colorlinks]{hyperref}  
\usepackage{amsmath,amssymb} 
\usepackage{array} 
\usepackage{multirow} 

\begin{document} 

\title{Cosmology in a supersymmetric model with gauged $B-L$}
\author{Anjishnu Sarkar} 
\email{anjishnu@phy.iitb.ac.in}
\author{Urjit A. Yajnik}
\email{yajnik@phy.iitb.ac.in}
\affiliation{Indian Institute of Technology, Bombay, Mumbai - 400076, India}
\date{}

\begin{abstract}
We consider salient cosmological features of a supersymmetric model which is 
left-right symmetric and therefore possessing gauged $B-L$ symmetry. The
requirement of breaking parity and also obtaining charge preserving vacua
introduces some unique features to this model (MSLRM), resulting in a 
preference for non-thermal leptogenesis. Assuming that the model preserves
TeV scale supersymmetry, we show that the vacuum structure generically
possesses domain walls, which can serve two important purposes. They can
signal a secondary inflation required to remove unwanted relics such as
gravitino and moduli and also generate lepton asymmetry by a mechanism
similar to electroweak baryogenesis. The requirement of disappearance of
domain walls imposes constraints on the soft parameters of the theory,
testable at the TeV scale. We also propose an alternative model
with spontaneous parity violation (MSLR\rlap/P). Incorporating the same
cosmological considerations in this case entails constraints on a different 
set of soft parameters. 
\end{abstract}
\maketitle

\section{Introduction}
\label{sec:intro} 

The minimal natural requirement that each of the three right handed 
neutrino states $\nu_{lR}$ be a doublet partner of the corresponding
right handed charged lepton $l_R$ results in a universal left-right 
symmetric gauge theory by demand, but also leads to gauging of an
important exact global symmetry of the standard model, viz., $B-L$. 
It also results in a satisfactory embedding for the electroweak 
hypercharge. A natural explanation of the very small observed neutrino 
masses then resides in the seesaw mechanism, with prediction of heavy
Majorana neutrino states, $N_i$, $i=1,2,3$, whose masses remain model 
dependent but substantially higher than the electroweak scale.  
While elegant, seesaw mechanism predicts a new high scale which gives rise 
to a hierarchy, further complicating the Higgs sector whose standard
model manifestation is also poorly understood. Inclusion of supersymmetry 
(SUSY) however improves the situation, stabilizing hierarchies of 
mass scales that lie above the SUSY breaking scale. We assume the
most optimistic value for the SUSY breaking scale, being the TeV
scale without disturbing the standard model. In this paper we study 
what has been called the minimal supersymmetric left-right symmetric
model (MSLRM) with the gauge group 
$SU(3)_c \otimes SU(2)_L \otimes SU(2)_R \otimes U(1)_{B-L}$. 
We explore the possibility for the left-right symmetric
scale to be low, a few orders of magnitude removed from the TeV scale. 
At a higher energy scale the model may turn out to be embedded in 
the supersymmetric $SO(10)$.

The most minimal supersymmetric left-right symmetric model considered  
\cite{km93, km95} fails to provide spontaneous breakdown of parity.
The addition of a parity odd singlet was considered in \cite{cve85}, however
this makes the charge preserving vacuum energetically disfavored \cite{km93}.
Aulakh et al. \cite{abs97, amrs98} discussed  the inclusion 
of two new Higgs triplet fields $\Omega$ and $\Omega_c$. In this model, 
$SU(2)_R$ first breaks to its subgroup $U(1)_R$, at a scale $M_R$, 
without affecting the $U(1)_{B-L}$. At a lower scale $M_{B-L}$,
$SU(3)_c \otimes SU(2)_L \otimes U(1)_R \otimes U(1)_{B-L}$ breaks to
$SU(3)_c \otimes SU(2)_L \otimes U(1)_Y$. In this scheme parity 
is spontaneously broken while preserving electromagnetic charge invariance.
However, due to parity invariance of the original theory, the 
phenomenologically unacceptable phase 
$SU(3)_c \otimes SU(2)_R \otimes U(1)_Y'$ is energetically degenerate with 
$SU(3)_c \otimes SU(2)_L \otimes U(1)_Y$.
An important consequence of this \cite{ys07} which we pursue here, is 
that  in the early  Universe, domain walls (DW) form at the scale $M_R$. 

Domain walls arising due to topological reasons \cite{kib80} play 
a crucial role in early cosmology.  If stable, they invalidate the model.
Independently, the over-abundance of the gravitino  and  moduli 
fields which are typically regenerated after the primordial 
inflation \cite{ekn84} is a generic problem in most SUSY models 
\cite{ekn84, ls96}. These two problematic ingredients of early cosmology 
have a happy bearing on each other. It has been shown \cite{mat00, kt05, ys07} 
that if DW dominate the evolution  of the universe  for a limited duration, 
the associated \textsl{secondary} inflation removes the gravitino and other 
dangerous relic fields like moduli. For this scenario to work,
it is crucial that the DW are metastable, with a decay temperature $T_D$ 
which must be larger than $\sim 10$ MeV in order to not interfere with big 
bang nucleosynthesis (BBN). While a scenario employing transient DW 
seems to lack the possibility of direct verification, it has been recently 
pointed out \cite{gs07} that the upcoming space based gravitational wave detectors 
may be able to detect the stochastic background arising at such phase
transitions.
In this paper we shall examine two models, MSLRM and MSLR{\rlap/P}, 
for the possibility of these requirements to be satisfied. The main result
is constraints on soft SUSY breaking parameters in the Higgs potentials.

Another issue of cosmology is that this class of models does not favor 
thermal leptogenesis for an intriguing  reason. $B-L$ asymmetry  in 
the form of fermion chemical potential 
is guaranteed to remain zero in the model until the gauged $B-L$ 
symmetry breaks spontaneously. As we shall see, a generic consequence of
the model is a relation among the various mass scales $M^2_{B-L} 
\simeq M_W M_R$, where 
$M_W$ is the electroweak scale. Thermal leptogenesis
requires $M_{B-L}$ to be larger than $10^{11}$-$10^{13}$ GeV, which pushes 
$M_R$ into the Planck scale in light of the above formula. A more 
optimistic constraint $M_{B-L} >10^9$GeV \cite{bbp03,bbp04}
requires left-right symmetry to be essentially grand unified theory.
More realistic scenarios therefore demand leptogenesis to be non-thermal 
in this class of models, either through bubble walls of a first 
order phase transition at the electroweak scale, or DW of the parity 
breaking phase transition.
It has been shown \cite{fglp91, sy05} that the only real requirement 
imposed by leptogenesis is that the presence of heavy neutrinos should not
erase lepton asymmetry generated by a given mechanism, possibly non-thermal. 
This places the modest bound $M_1> 10^4$GeV, on the mass of the lightest 
of the  heavy Majorana neutrinos. 

The paper is organized as follows. We first review the MSLRM in sec. 
\ref{sec:review}. We then outline the main features of cosmology in such a 
model in sec. \ref{sec:sybr-cosmo}.
In sec. \ref{sec:pslash} we introduce the new model MSLR\rlap/P and discuss
the differences to cosmology that can arise. In sec. 
\ref{sec:gravitino-dilution} we
identify the condition for gravitino dilution and its consistency with
the $M_R$ scale acceptable in these models.  In secs. 
\ref{sec:remove-mslrm} and \ref{sec:remove-pslash} we write down the permissible
soft terms in the two models and obtain the constraints on the parameters
determining the safe disappearance of domain walls. Sec. \ref{sec:conclusion}
contains summary of conclusions and an outlook.

\section{Degenerate vacua of MSLRM} 
\label{sec:review} 

The minimal supersymmetric left-right symmetric model (MSLRM) \cite{abs97}
contains quark and lepton superfields, one set for each generation, with 
their quantum numbers  under $SU(3)_c$, $SU(2)_L$, $SU(2)_R$, $U(1)_{B-L}$ respectively 
given by
\begin{eqnarray}
Q = (3,2,1,1/3), & \quad & Q_c = (3^*,1,2,-1/3), \nonumber  \\
L = (1,2,1,-1),  & \quad & L_c = (1,1,2,1).
\end{eqnarray}
where we have suppressed the generation index. The minimal set of Higgs 
superfields required is,
\begin{eqnarray} 
\Phi_i = (1,2,2,0),    & \quad & i = 1,2, \nonumber \\
\Delta = (1,3,1,2),    & \quad & \bar{\Delta} = (1,3,1,-2), \nonumber \\
\Delta_c = (1,1,3,-2), & \quad & \bar{\Delta}_c = (1,1,3,2),
\end{eqnarray}
where the bidoublet is doubled so that the model has non-vanishing 
Cabibo-Kobayashi-Maskawa matrix. The number of triplets is doubled 
to have anomaly cancellation.
Under discrete parity symmetry the fields are prescribed to transform as,
\begin{eqnarray}
Q \leftrightarrow Q_c^*, \quad & 
L \leftrightarrow L_c^*, \quad & 
\Phi_i \leftrightarrow \Phi^\dagger_i,  \nonumber \\
\Delta \leftrightarrow \Delta_c^*,  \quad & 
\bar{\Delta} \leftrightarrow \bar{\Delta}_c^*.
\label{eq:parity} 
\end{eqnarray}
It has been demonstrated in \cite{km93} that this minimal generalization
to supersymmetry results in a vacuum with $\langle \Delta \rangle =$
$\langle \bar{\Delta} \rangle =$ 
$\langle \Delta_c \rangle =$ 
$\langle \bar{\Delta_c} \rangle $.
Thus the electroweak scale physics would no longer be chiral.
This was also demonstrated to be a general feature for a class of
related models.  It was proposed to cure this problem by introducing
a parity odd singlet \cite{cve85}.
However this results in electromagnetic charge violating vacua.
This conclusion in turn can be avoided, but at the cost of violating 
$R$ parity.

These problems were circumvented in \cite{abs97}, i.e., spontaneous 
parity breaking, preserving electromagnetic charge invariance, 
and retaining $R$ parity, can all be achieved by introducing two new 
triplet Higgs fields with the  following charges.
\begin{equation} 
\Omega = (1,3,1,0), \qquad \Omega_c = (1,1,3,0) ~. 
\end{equation} 
Under parity $\Omega \leftrightarrow \Omega_c^*$.
We shall assume that supersymmetry is broken only at the electroweak scale.
Thus at higher scales we look for supersymmetry preserving vacua. Such vacua 
of the theory are obtained by imposing F-flatness and  D-flatness conditions.
The conditions can be found in \cite{abs97} and are similar to the case we
have worked out in the Appendix \ref{sec:appendix} for the modified version 
of this model we propose later in sec. \ref{sec:pslash}. 
The conditions lead to the following set of vaccum expectation value (vev's) 
for the Higgs fields as one of the possibilities,
\begin{equation} 
\begin{array}{ccc}
\langle \Omega_c \rangle = 
\begin{pmatrix}
\omega_c & 0 \\
0 & - \omega_c
\end{pmatrix}, & \qquad
\langle \Delta_c \rangle =
\begin{pmatrix}
0 & 0 \\
d_c & 0
\end{pmatrix}, & \qquad
\langle \bar{\Delta}_c \rangle =
\begin{pmatrix}
0 & \bar{d}_c \\
0 & 0
\end{pmatrix}, \\ [0.5cm] 
\langle \Omega \rangle = 0, & \qquad
\langle \Delta \rangle = 0, & \qquad
\langle \bar{\Delta} \rangle = 0.
\label{eq:rhvev} 
\end{array}
\end{equation}
At the scale $M_R$, when $\Omega$, $\Omega_c$ acquire vev, $SU(2)_R$ 
is broken to $U(1)_R$. At a lower scale $M_{B-L}$, the vev of the triplet 
Higgs fields breaks $SU(3)_c \otimes SU(2)_L \otimes$ 
$U(1)_R \otimes U(1)_{B-L}$ to $SU(3)_c \otimes SU(2)_L \otimes$ $U(1)_Y$. 
Thus, at TeV scale, the model breaks exactly to the minimal supersymmetric 
standard model (MSSM). It was shown \cite{abs97} that this scheme of 
breaking preserved electromagnetic charge invariance and parity was 
spontaneously broken. Further, although the $\Delta$ fields signal $B-L$
breaking, $R$ parity  is preserved due to the fact that 
\begin{equation}
R=(-1)^{3(B-L)+2s}
\end{equation}
and the $\Delta$ field vev's violate $B-L$ by at least $2$ units.

We now observe that the $D$ and $F$ flatness conditions imposed above also 
permit a vacuum preserving the $SU(2)_R \otimes U(1)_L \otimes U(1)_{B-L}$ 
symmetry which is energetically degenerate with the one identified in 
eqns. (\ref{eq:rhvev}). The alternative set of vev's is given by
\begin{equation} 
\begin{array}{ccc}
\langle \Omega_c \rangle = 0, & \qquad
\langle \Delta_c \rangle = 0, & \qquad
\langle \bar{\Delta}_c \rangle = 0, \\[0.25cm]
\langle \Omega \rangle = 
\begin{pmatrix}
\omega & 0 \\
0 & - \omega
\end{pmatrix}, & \qquad
\langle \Delta \rangle =
\begin{pmatrix}
0 & 0 \\
d & 0
\end{pmatrix}, & \qquad
\langle \bar{\Delta} \rangle =
\begin{pmatrix}
0 & \bar{d} \\
0 & 0
\end{pmatrix}.
\label{eq:lhvev} 
\end{array}
\end{equation}
This is only to be expected from the $L\leftrightarrow R $ symmetry
of the model. As we discuss below this leads to the formation of
domain walls and we must have a mechanism in the model for the
removal of such walls.

\section{Pattern of symmetry breaking, cosmology and baryogenesis}
\label{sec:sybr-cosmo}
The solutions for the vev's from the conditions (\ref{eq:rhvev}) are
\begin{eqnarray} 
\omega &=& -\frac{m_\Delta}{a}\equiv -M_R, \nonumber \\
d=\bar{d} &=&
\left( \frac{2m_\Delta m_\Omega}{a^2}\right)^{1/2} \equiv M_{B-L}
\label{eq:vevvalues}
\end{eqnarray} 
As discussed earlier the requirement of obtaining parity breakdown 
is $M_R \gg M_{B-L}$, which can be arranged by choosing $m_\Delta \gg m_\Omega$
which leads to the relation $M_{B-L}^2 \simeq M_R M_\Omega$. We also
adopt the proposal of \cite{amrs98} and assume that $m_\Omega \simeq m_W$.
This is natural, provided the $m_\Omega$ originates from the soft 
SUSY breaking sector, which also means in turn that $m_\Omega$ is 
comparable to gravitino mass $m_{3/2}$. Thus we effectively have the
relation 
\begin{equation}
M_{B-L}^2 \simeq M_R M_W. 
\label{eq:scale-rel} 
\end{equation}

The most optimistic value for $M_{B-L}$ is $\sim 10^4 $ GeV with
corresponding $M_R \sim 10^6$ GeV. On the other hand, the largest
value of $M_R$ is the intermediate scale value $\sim \sqrt{M_{Pl}M_W}$
$\sim 10^{10}$ GeV,
beyond which non-renormalizable Planck scale corrections begin to 
be significant. This corresponds to $M_{B-L}\sim 10^6$ GeV.
We shall thus be interested in this range of values for $M_R$ and
$M_{B-L}$, however the lower values make the model amenable to investigation
at colliders.

The model contains DW separating the phases identified in (\ref{eq:rhvev}) 
and (\ref{eq:lhvev}). The DW must be unstable for cosmological reasons. 
The required asymmetry between the two types of vacua has to arise 
dynamically. Since this is not admissible in the superpotential, it must 
arise from the soft terms \cite{ys07}. This means that the mechanism 
inducing the soft terms must cause a bias between the two types of vacua. In
a gauge mediated supersymmetry breaking (GMSB) scenario, the hidden sector 
or the messenger sector or both must cause a distinction between 
the two kinds of vacua. On the other hand in a gravity mediated
scenario it is permissible to violate the discrete symmetry due to
gravitational effects.

An alternative possibility for parity breaking is discussed in
section \ref{sec:pslash}. In both cases, below the TeV scale the theory 
is effectively MSSM. Due to the
coupling of the bidoublets $\Phi_i$ to $\Omega_c$, one pair of 
doublets becomes heavy and only one pair of $SU(2)_L$ doublets 
remains light. 

The discussion so far shows that
the model has a characteristic cosmological history, with
the $SU(2)_R$ breaking first when the $\Omega$ fields acquire a vev.
This is accompanied by the formation of
domain walls, with SUSY still preserved. The DW come to dominate the
energy density of the Universe and cause the onset of secondary inflation.
As the temperature reduces, at the scale $M_{B-L}$, the triplets
$\Delta$ acquire vev and the heavy neutrinos obtain Majorana mass. 
In principle the DW persist down to electroweak scale
and lower, possibly upto a decay temperature $T_D$ in the range 
$10$ MeV-$10$ GeV \cite{kt05}. The secondary inflation helps to remove 
unwanted relics viz., gravitinos and moduli fields regenerated after the GUT 
or Planck scale inflation. 

The final disappearance of DW completes the phase transition which commenced 
at $M_R$. We designate this temperature as $T_D$.
The decay process of DW gives rise to entropy production and reheating
in a model dependent way.
In several models the energy scale of the first order phase transition
can be such as to leave behind stochastic gravitational wave
background detectable at upcoming space based
gravitational wave experiments as pointed out by \cite{gs07}.

Since both $B$ and $L$ are effectively conserved below the electroweak
scale, baryogenesis requires that the reheat temperature should be 
significantly higher than that of the electroweak phase transition.
The reheat temperature after secondary inflation $(T^s_R)$ however, should 
not be so high as to regenerate the unwanted relics. Thus, in this model
it is required that the secondary inflation has $10^9 > $ 
$T^s_R$ $ > 10^2$ GeV. The upper limit on 
$T^s_R$ is seen to be easily satisfied
for most of the range of values for $M_R \sim 10^6 - 10^{10}$ GeV 
in this class of models as seen from the energy scales of symmetry breaking.

Furthermore, we saw that the natural scale of $M_{B-L}$ is $10^4\,-\,10^6$GeV. 
Thus thermal leptogenesis is also disfavored and must proceed
via one of several possible non-thermal mechanisms.
Low scale leptogenesis with special attention to left-right symmetry
and supersymmetric mechanisms has already received attention in several 
works \cite{cynr02, bou02, ham02, bhs04,  cs05, sco06, mprs07}.

Specifically the conditions outlined in \cite{cynr02} are easily
seen to be satisfied in the present class of models.
(i) The phase transition at the $M_R$ scale is necessarily first order 
due to formation of DW in turn inevitable due to parity 
invariance of the underlying theory. This ensures the out-of-equilibrium 
condition necessary for leptogenesis.
(ii) Since the DW decay must return the universe to 
$SU(2)_L$ preserving rather than $SU(2)_R$ preserving phase their is a 
directionality to DW motion resulting in time reversal violation.
(iii) The presence of several complex scalar fields in the model allows the 
formation of CP violating condensate in the core of the DW.
(iv) This CP violating phase can enter the Dirac mass matrix of the 
neutrinos streaming through the DW giving rise to leptogenesis governed by 
the ``classical force" mechanism \cite{jpt95, jpt_etal_96b, ck00, cjk98, 
cynr02}. 
(v) The lepton asymmetry gets partially converted to baryon asymmetry
due to the action of the sphalerons.

\section{Supersymmetric left-right model with spontaneous parity breaking 
(MSLR\rlap/P)}
\label{sec:pslash}
We have argued above the need for GMSB or gravity induced soft terms 
in MSLRM that can lift 
the symmetry between left and right vacua. While it is easy to see
that only one of the two phases can survive, understanding of the 
selection of the observed vacuum requires additional details in  
the Planck scale model.

A more appealing option has been considered by  Chang et al. 
\cite{cmp84} where spontaneous breaking of parity is implemented
within the Higgs structure of the theory. The model is based on 
$SU(3)_c \otimes SU(2)_L \otimes SU(2)_R \otimes U(1)_{B-L} \otimes P$,
where $P$ denotes a parity symmetry. A gauge singlet field $\eta$ is
introduced which is odd under $P$, viz., $\eta\leftrightarrow -\eta$.
The potential of the Higgs fields contains a term
\begin{equation}
V_{\eta\Delta} \sim M \eta (\Delta^\dagger_L \Delta_L - \Delta^\dagger_R \Delta_R )
\end{equation}
so that if $\eta$ acquires a vev at a high scale $M_P \gg M_W$, 
$SU(2)_R$ is not broken, but the effective 
masses of the $\Delta_L$ and $\Delta_R$ become different and the 
$ L\leftrightarrow R$ symmetry appears explicitly broken.

A direct implementation of this idea in supersymmetric theory however, would 
lead us back to the model of Kuchimanchi and Mohapatra \cite{km93} and 
charge breaking vacua. We propose an alternative model 
based on the group 
$SU(3)_c \otimes SU(2)_L \otimes SU(2)_R \otimes U(1)_{B-L} \otimes P$,
with a pair of triplets as in \cite{abs97}. 
However, unlike \cite{abs97} the Higgs triplets $\Omega$, $\Omega_c$ 
are odd under the parity symmetry. Specifically, 
\begin{eqnarray}
Q \leftrightarrow Q_c, \quad & 
L \leftrightarrow L_c, \quad & 
\Phi_i \leftrightarrow \Phi^T_i,  \nonumber \\
\Delta \leftrightarrow \Delta_c,  \quad & 
\bar{\Delta} \leftrightarrow \bar{\Delta}_c, \quad & 
\Omega \leftrightarrow -\Omega_c.
\label{eq:imparity}
\end{eqnarray}
We dub this model MSLR{\rlap/}P. The superpotential consistent with
this parity is given by the following expression, with a few essential 
differences from the superpotential of \cite{abs97}.
\begin{eqnarray}
 W_{LR}&=& {\bf h}_l^{(i)} L^T \tau_2 \Phi_i \tau_2 L_c 
 + {\bf h}_q^{(i)} Q^T \tau_2 \Phi_i \tau_2 Q_c 
 +i {\bf f} L^T \tau_2 \Delta L 
 +i {\bf f} L^{cT}\tau_2 \Delta_c L_c \nonumber \\
&& + ~m_\Delta  {\rm  Tr}\, \Delta \bar{\Delta} 
  + m_\Delta  {\rm Tr}\,\Delta_c \bar{\Delta}_c
  + \frac{m_\Omega}{2} {\rm Tr}\,\Omega^2 
  + \frac{m_\Omega}{2} {\rm Tr}\,\Omega_c^2 \nonumber \\
&& + ~\mu_{ij} {\rm Tr}\,  \tau_2 \Phi^T_i \tau_2 \Phi_j 
  +a {\rm Tr}\,\Delta \Omega \bar{\Delta}
  -a {\rm Tr}\,\Delta_c \Omega_c \bar{\Delta}_c \nonumber \\
& &  + ~\alpha_{ij} {\rm Tr}\, \Omega  \Phi_i \tau_2 \Phi_j^T \tau_2 
  - \alpha_{ij} {\rm Tr}\, \Omega_c  \Phi^T_i \tau_2 \Phi_j \tau_2 ~,
\label{eq:imsuperpot}
\end{eqnarray} 
where color and flavor indices have been suppressed.
Further, ${\bf h}^{(i)}_{q}  =  {{\bf h}^{(i)}_{q}}^\dagger $,
${\bf h}^{(i)}_{l}  =  {{\bf h}^{(i)}_{l}}^\dagger $,  
$\mu_{ij}  =  \mu_{ji} = \mu_{ij}^*$,
$\alpha_{ij} = -\alpha_{ji}$. Finally, ${\bf f}$, ${\bf h}$ 
are real symmetric matrices with respect to flavor indices. 

The $F$ and $D$ flatness conditions derived from this superpotential are 
presented in appendix \ref{sec:appendix}. However, the effective
potential for the scalar fields which is determined from modulus
square of the $D$ terms remains the same as for the MSLRM at least 
for the form of the ansatz of the vev's we have chosen. 
As such the resulting solution for the vev's remains identical 
to eq. (\ref{eq:vevvalues}). The difference in the effective
potential shows up in the soft terms as will be shown later. 
Due to soft terms, below  the scale $M_R$ the effective 
mass contributions to $\Delta_c$ and $\bar{\Delta}_c$ become
larger than those of $\Delta$ and $\bar{\Delta}$. The cosmological
consequence of this is manifested after the $M_{B-L}$ phase transition
when the $\Delta$'s become massive. Unlike MSLRM where the DW 
are destabilized only after the soft terms become significant, i.e., 
at the electroweak scale, the DW in this case become unstable immediately
after $M_{B-L}$. Leptogenesis therefore commences immediately below 
this scale and the scenario becomes qualitatively different from 
that for the MSLRM. This is a subject for future study.
In the present work we focus on the removal of unwanted relics 
and safe exit from DW dominated secondary inflation. 

\section{Dilution of Gravitinos} 
\label{sec:gravitino-dilution}

It is reasonable to assume that any primordial abundance of 
gravitinos  has been diluted by the primordial inflation. 
The gravitinos with potential consequences to observable cosmology are 
generated entirely after reheating of the universe 
($T_R \sim 10^9 \textrm{GeV}$) subsequent to primordial inflation.
Detailed calculations  \cite{ekn84}  show that the gravitino number 
density ($n_{3/2}$) at a low temp $T_f$ is given by
\begin{equation} 
n_{3/2}(T_f) = 3.35 \times 10^{-12} ~T_9^{\textrm{max}} ~T_f^3 
\times (1-0.018 \ln T_9^{\textrm{max}}) ~,
\end{equation} 
where, $T^{\textrm{max}}_9 = T/ 10^{9} \textrm{~GeV}$

Here we  constrain the dynamics governing the DW by 
requiring that they dilute this regenerated gravitino abundance 
adequately. The possible values of $M_R$ and $M_{B-L}$ are discussed
below eq. (\ref{eq:scale-rel}). Here onwards we shall assume $M_R=10^6$GeV 
and $M_{B-L}=10^4$GeV as an example.
Putting $T_f \approx M_R = 10^{6} \textrm{~GeV}$ in the above eqn. 
we get,
\begin{equation} 
n_{3/2}(T_f) = 3.35 \times 10^6 (\textrm GeV)^3 \equiv n_{3/2}^{b} ~,
\label{eq:nthreehalf}
\end{equation} 
where, $n_{3/2}^{b}$ is the gravitino number density at the beginning of 
secondary inflation at the temperature $T_f = 10^6 \textrm{GeV}$.
The number density of gravitino $(n_{3/2})$, during this time, decreases 
as $R^{-3}$, where $R$ is the scale factor. Therefore we have,
\begin{equation} 
R_{e} = R_{b} \left(\frac{n_{3/2}^{b}}{n_{3/2}^{e}}\right)^{1/3} ~,
\label{eq:g_dil} 
\end{equation} 
where, $R_{b}$ ($R_{e}$) and $n_{3/2}^{b}$ ($n_{3/2}^{e}$)
are the scale factor and gravitino density
at beginning (end) of secondary inflation. 

The best constraint that can be imposed on the gravitinos, produced after 
primordial inflation, comes from the fact that entropy produced due to decay 
of gravitino shouldn't disturb the delicate balance of light nuclei 
abundance \cite{lin80, ekn84}. This constraint is given by,
\begin{equation} 
\frac{m_{3/2} f \beta n_{3/2}}{n_e E_{\star}} \lesssim 1 ~. 
\end{equation} 
Here, $f$ is the fraction of the gravitino number density $(n_{3/2})$
that dumps its entropy in the universe, and is taken to be $f = 0.8$ \cite{ekn84},
$m_{3/2} = 100$ GeV is the mass of the gravitino and
$T$ is the scale of BBN taken to be $\sim 1$MeV.
Finally, $E_{\star} = 100 \textrm{ MeV}$ \cite{lin80}, and 
$\beta$  is a mildly temperature dependent parameter, with
numerically determined value $1.6$, which lead to the estimate
\begin{equation} 
n_{3/2} \lesssim 1.66 ~\delta_B \times 10^{-13} (\textrm{GeV})^3 
\equiv n_{3/2}^{e} ~, 
\end{equation} 
where $\delta_B \left[\equiv n_B/n_\gamma \right]$ is the baryon to photon 
ratio and $n_\gamma \left[= (2 \zeta(3)/\pi^2) ~T^3 \right]$ is the photon 
number density.
Using the Wilkinson Microwave Anisotropic Probe (WMAP) data
$\left( \delta_B = \left(6.1^{+0.3}_{-0.2}\right) \times 10^{-10} \right)$ 
\cite{benn_etal}
We find that
\begin{equation}
 n^{e}_{3/2} = 1.013 \times 10^{-22} (\textrm{GeV})^3 ~. 
\end{equation}
So from eq (\ref{eq:g_dil}) we have,
\begin{equation}
R_{e} = R_{b} \times 3.2 \times 10^9 ~.
\end{equation}
Therefore, number of {\it e} foldings required is
\begin{equation}
 N_e =  \ln \left( \frac{R_{e}}{R_{b}} \right)
= \ln 10^{9} \backsimeq 20  ~. 
\label{eq:efolding}
\end{equation}
This agrees with the observation by \cite{mat00} that a secondary
inflation can dilute the moduli and gravitinos sufficiently that
no problem results for cosmology. As pointed out by \cite{ls96} there can also 
be more than one secondary inflation to effectively reduce the 
moduli/gravitino number density. Here we shall assume this to the only
secondary inflation sufficient for diluting the gravitino density.

Finally, a handle on the explicit symmetry breaking parameters 
of the two models can be obtained by noting that there should exist
sufficient wall tension for the walls to disappear before
a desirable temperature scale $T_D$. It has been observed in
\cite{ptww91} that the free energy density difference $\delta \rho$
between the vacua, which determines the pressure difference across 
a domain wall should be of the order 
\begin{equation}
\delta \rho \sim T_D^4
\label{eq:dr_Td_rel}
\end{equation}
in order for the DW structure to disappear at the scale $T_D$. 

In the scenarios we consider in the next two sections,
the DW form at the higher scale $M_R$. However at this
point the thermal vacua on the two sides of the walls are both
equal in free energy. At a lower scale, due to additional
symmetry breaking taking effect, more field condensates 
become a part of the DW structure. After such changes
the free energy balance across the walls can change and
a net $\delta \rho$ can arise. We shall now see specifically
the sources of such changes and relate the resulting
$\delta \rho$ to the parameters in appropriate potentials.

\section{Removal of domain walls : MSLRM}
\label{sec:remove-mslrm}
The possible source for breaking the parity symmetry of the
MSLRM lies in soft terms with the assumption that the 
hidden sector, or in case of GMSB also
perhaps the messenger sector does not obey the parity of the
visible sector model. For gravity mediated breaking this can
be achieved in a natural way since a discrete symmetry can be 
generically broken by gravity effects. We present the possible
soft terms for MSLRM below.
\begin{eqnarray} 
\mathcal{L}_{soft} &=& 
\alpha_1 \textrm{Tr} (\Delta \Omega \Delta^{\dagger}) +
\alpha_2 \textrm{Tr} (\bar{\Delta} \Omega \bar{\Delta}^{\dagger}) +
\alpha_3 \textrm{Tr} (\Delta_c \Omega_c \Delta^{\dagger}_c) + 
\alpha_4 \textrm{Tr} (\bar{\Delta}_c \Omega_c \bar{\Delta}^{\dagger}_c) ~~~~~
\label{eq:sigNdel} \\ 
&& + ~m_1^2 \textrm{Tr} (\Delta \Delta^{\dagger}) +
m_2^2 \textrm{Tr} (\bar{\Delta} \bar{\Delta}^{\dagger}) + 
m_3^2 \textrm{Tr} (\Delta_c \Delta^{\dagger}_c) +
m_4^2 \textrm{Tr} (\bar{\Delta}_c \bar{\Delta}^{\dagger}_c) 
\label{eq:delta} \\
&& + ~\beta_1 \textrm{Tr} (\Omega \Omega^{\dagger}) +
\beta_2 \textrm{Tr} (\Omega_c \Omega^{\dagger}_c) ~.
\label{eq:omega} 
\end{eqnarray} 
The contributions to the free energy difference $\delta \rho$ 
i.e. difference between the left and right sector, 
can now be estimated from the above Lagrangian.
It is natural to consider $\alpha_1 \sim \alpha_2$ and $\alpha_3 \sim \alpha_4$.
In this case it can be shown that the use of eq. (\ref{eq:dr_Td_rel}) 
does not  place a severe constraint on the $\alpha_i$'s 
For the rest of the soft terms [(\ref{eq:delta}) and (\ref{eq:omega})]
we have respectively, in obvious notation
\begin{equation}
{\delta \rho}_\Delta = 
\left[ m_1^2 \textrm{Tr} (\Delta \Delta^{\dagger}) +
m_2^2 \textrm{Tr} (\bar{\Delta} \bar{\Delta}^{\dagger}) \right] 
- \left[ m_3^2 \textrm{Tr} (\Delta_c \Delta^{\dagger}_c) +
m_4^2 \textrm{Tr} (\bar{\Delta}_c \bar{\Delta}^{\dagger}_c) \right] 
= 2(m^2 - m^{2\prime}) d^2 ~,
\label{eq:ep_delta}  
\end{equation}
\begin{equation}
{\delta \rho}_\Omega = \beta_1 \textrm{Tr} (\Omega \Omega^{\dagger}) -
\beta_2 \textrm{Tr} (\Omega_c \Omega^{\dagger}_c)  
= 2(\beta_1 - \beta_2) ~\omega^2 ~,
\label{eq:ep_omega} 
\end{equation}
where we have considered 
$m_1^2 \backsimeq m_2^2 \equiv m^2$, $m_3^2 \backsimeq m_4^2 \equiv 
m^{\prime~2}$.
The vev's of neutral component of $\Delta (\Delta_c)$ and $\Omega (\Omega_c)$
are $d (d_c)$ and $\omega (\omega_c)$. Here we have assumed that
$d_c \sim d$ and $\omega_c \sim \omega$.

Using the constraint (\ref{eq:dr_Td_rel}) in the eqns. 
(\ref{eq:ep_delta}), (\ref{eq:ep_omega}), 
we can determine the differences between the relevant soft parameters
for a range of permissible values of $T_D$. 
\begin{table}
\begin{center} 
{\setlength\extrarowheight{1.5mm}
\begin{tabular}{c|c|c|c} 
\hline
& $T_D = 10$ GeV & $T_D = 10^2$ GeV & $T_D = 10^3$ GeV \\ \hline \hline
$(m^2 - m^{2\prime}) \sim$ & 
$10^{-4}\textrm{ GeV}^2$ & $1 \textrm{ GeV}^2$ & $10^{4} 
\textrm{ GeV}^2$\\ [1mm]
$(\beta_1 - \beta_2) \sim$ & 
$10^{-8}\textrm{ GeV}^2$ & $10^{-4}\textrm{ GeV}^2$ & 
$1 \textrm{ GeV}^2$ \\ [1mm] \hline \hline
\end{tabular} }
\end{center}
\caption{Asymmetry in parameters, for a range of $T_D$, signifying 
magnitude of explicit parity breaking}
\label{tab:DWalls}
\end{table}
In Table \ref{tab:DWalls} we have taken $d \sim 10^4 $ GeV, 
$\omega \sim 10^6 $ GeV and $T_D$ in the range 
$100 \textrm{ MeV} - 10 \textrm{ GeV}$ \cite{kt05}.
The above differences between the values in the left and right sectors is a 
lower bound on the soft parameters and is very small. Larger values would
be acceptable to low energy phenomenology. However if we wish to retain the
connection to the hidden sector, and have the advantage of secondary 
inflation we would want the differences to be close to this bound.
As pointed out in \cite{ptww91, dn93} an asymmetry $\sim 10^{-12}$ 
is sufficient to lift the degeneracy between the two sectors.

\section{Removal of domain walls : MSLR\rlap/P}
\label{sec:remove-pslash}
In this model parity breaking is achieved spontaneously within
the observable sector below the scale $M_R$ at which the $\Omega$
fields acquire vev's. However the breaking is not manifested in the
vacuum till the scale $M_{B-L}$ where the $\Delta$ fields acquire 
vev's. For simplicity 
we assume that the hidden sector  responsible for SUSY breaking
does not contribute parity breaking terms. This is reasonable
since even if the hidden sector breaks this parity the corresponding 
effects are suppressed by the higher scale of breaking
and in the visible sector the parity breaking effects are
dominated by the explicit mechanism proposed. Thus at a scale
above $M_R$ but at which SUSY is broken 
in the hidden sector we get induced soft terms respecting this
parity. Accordingly, for the Higgs sector the parameters can be 
chosen such that 
\begin{eqnarray} 
\mathcal{L}_{soft} &=& 
\alpha_1 \textrm{Tr} (\Delta \Omega \Delta^{\dagger}) -
\alpha_2 \textrm{Tr} (\bar{\Delta} \Omega \bar{\Delta}^{\dagger}) -
\alpha_1 \textrm{Tr} (\Delta_c \Omega_c \Delta^{\dagger}_c) +
\alpha_2 \textrm{Tr} (\bar{\Delta}_c \Omega_c \bar{\Delta}^{\dagger}_c) ~~~~~
\label{eq:imsigNdel} 
\nonumber \\ 
&& + ~m_1^2 \textrm{Tr} (\Delta \Delta^{\dagger}) +
m_2^2 \textrm{Tr} (\bar{\Delta} \bar{\Delta}^{\dagger}) + 
m_1^2 \textrm{Tr} (\Delta_c \Delta^{\dagger}_c) +
m_2^2 \textrm{Tr} (\bar{\Delta}_c \bar{\Delta}^{\dagger}_c) 
\label{eq:imdelta} 
\nonumber \\
&& + ~\beta \textrm{Tr} (\Omega \Omega^{\dagger}) +
\beta \textrm{Tr} (\Omega_c \Omega^{\dagger}_c) ~.
\label{eq:imomega} 
\end{eqnarray} 
These terms remain unimportant at first due to the key assumption 
leading to MSSM as the effective low energy theory. The SUSY
breaking effects become significant only at the electroweak scale.
However, below the scale $M_R$, $\Omega$ and $\Omega_c$ acquire vev's
given by eq. (\ref{eq:rhvev}) or (\ref{eq:lhvev}).
Further, below the scale $M_{B-L}$ the $\Delta$ fields acquire vev's
and become massive. The combined contribution from the superpotential 
and the soft terms to the $\Delta$ masses now explicitly encodes the
parity breaking,
\begin{equation} 
\begin{array}{cc}
\mu^2_\Delta = M^2_\Delta + \alpha_1 \omega, & \qquad 
\mu^2_{\Delta_c} = M^2_\Delta - \alpha_1 \omega, \\ [3mm]
\mu^2_{\bar{\Delta}} = M^2_\Delta + \alpha_2 \omega,  & \qquad 
\mu^2_{\bar{\Delta}_c} = M^2_\Delta - \alpha_2 \omega. 
\end{array}
\label{eq:thiggs_mc} 
\end{equation} 
where $M^2_\Delta$ is the common contribution from the superpotential.
The difference in free energy across the domain wall is now 
dominated by the differential contribution to the $\Delta$ masses
\begin{equation} 
\delta\rho_\alpha \equiv 2(\alpha_1 + \alpha_2) \omega d^2, 
\label{eq:lrasymm} 
\end{equation} 
where we have considered 
$\omega_c \sim \omega$, $d \sim \bar{d} \sim d_c \sim \bar{d}_c$.
Now using  eq (\ref{eq:dr_Td_rel}) for a range of temperatures 
$(T_D \sim 10^2 \text{ GeV} - 10^4 \text{ GeV})$,
determines the corresponding range of values
of coupling constants as 
\begin{equation} 
(\alpha_1 + \alpha_2) \sim 10^{-6} - 10^{2} \text{ GeV},
\label{eq:a_diff} 
\end{equation} 
where we have considered $|\omega| \simeq M_R$, $|d| \simeq M_{B-L}$.

\section{Conclusions}  
\label{sec:conclusion} 

Supersymmetric left-right model is an appealing model to consider
beyond the MSSM due to its natural inclusion of right handed neutrino
and gauged $B-L$ symmetry. There is a generic problem with building
this kind of models due to their inability to preserve electromagnetic
charge invariance together with preserving $R$ parity. MSLRM solves
the problem by breaking $SU(2)_R$ at a higher scale and $U(1)_{B-L}$ 
at a lower scale. A generic consequence of this improvement is a relation 
$M_{B-L}^2\simeq M_RM_W$. If $M_R$ is at most of intermediate scale value
$10^{10}$GeV we get the natural range of values $10^4$GeV$\,-\,10^6$GeV
for the $M_{B-L}$. 
A generic problem of this class of models remains the need to select
the $SU(2)_L$ as the low energy gauge group as against $SU(2)_R$.
We propose a new model MSLR\rlap/P with spontaneous parity breaking arising
in Higgs sector. Both models face the same problems as MSSM with regard to
baryogenesis and there is a strong case for leptogenesis to 
be non-thermal due to the low natural scale for $M_{B-L}$.
The pattern of breaking and associated cosmological events in
the two classes of models are summarized in table \ref{tab:pattern}.

\begin{table}[!htp]
\begin{center}
\setlength{\extrarowheight}{1.5mm} 
\begin{tabular}{c|c|c|c|c}
\hline
Cosmology & Scale & Symmetry Group & MSLR\rlap/P & MSLRM \\
&&&(GeV)&(GeV) \\ [1.5mm] \hline \hline
\multirow{3}{1.75in}{%
$\Omega$ or $\Omega_c$ get vev. \\
Onset of wall dominated \\ secondary inflation.} & &
$SU(3)_c \otimes SU(2)_L \otimes SU(2)_R \otimes U(1)_{B-L}$ & &
\\
& $M_R$ & $\downarrow$ & $10^6$ & $10^6$  \\
&&&& 
\\ [1.5mm] \hline
\multirow{2}{1.75in}{%
Higgs triplet $(\Delta's)$ \\ get vev } & &
$SU(3)_c \otimes SU(2)_L \otimes U(1)_R \otimes U(1)_{B-L}$ && \\
& $M_{B-L}$ & $\downarrow$ & $10^4$ & $10^4$ \\ [1.5mm] \hline
\multirow{2}{1.75in}{%
End of inflation and \\ beginning of L-genesis} &
$M_{B-L}$
&& $10^4$ & --- \\ \cline{2-5} 
& $M_S$ & & --- & $10^3$ \\ [1.5mm] \hline
\multirow{2}{*}{SUSY breaking} & & 
$SU(3)_c \otimes SU(2)_L \otimes U(1)_Y$ (SUSY) & & \\ 
& $M_S$ & $\downarrow$ & $10^3$ & $10^3$ \\ [1.5mm] \hline
\multirow{2}{1.75in}{Wall disapperance \\ temperature} & 
\multirow{2}{*}{$T_D$} & & 
\multirow{2}{*}{$10 - 10^3$} & 
\multirow{2}{*}{$10 - 10^2$} \\
&&&& \\ [1.5mm] \hline
\multirow{2}{1.75in}{Secondary reheat \\ temperature} & 
\multirow{2}{*}{$T^s_R$} & & 
\multirow{2}{*}{$10^3 - 10^4$} & 
\multirow{2}{*}{$10^3$} \\
&&&&\\[1.5mm] \hline
\multirow{2}{*}{Electroweak breaking} & &
$SU(3)_c \otimes SU(2)_L \otimes U(1)_Y$ (non-SUSY)& & 
\\ & $M_W$ & $\downarrow$ & $10^2$ & $10^2$ \\[1.5mm] \hline
Standard model & & $SU(3)_c \otimes U(1)_{EW}$ & & 
\\ [1.5mm] \hline \hline
\end{tabular}
\caption{Pattern of symmetry breaking and the slightly different sequence
of associated cosmological events in the two classes of models}
\label{tab:pattern}
\end{center} 
\end{table} 

Parity invariance of the underlying theory gives rise to domain walls in the early
Universe. While this is problematic if the walls are stable, it can be
shown that a low energy ( $T< 10^9$GeV) epoch dominated by transient  
domain walls can help to remove unwanted relics generic to supersymmetry. 
In eq.s (\ref{eq:nthreehalf}) and (\ref{eq:efolding}) we obtain a requirement on 
the extent of secondary inflation caused by the domain walls, relating it to the 
energy scale $M_R$.  
We explored permissible values of $T_D$, ( the temperature scale at which the 
domain walls finally disappear) for the walls to be unstable
despite causing sufficient secondary inflation.
The conditions appear as limits on differences of parameters in the 
supersymmetry  breaking soft terms. The values in Table \ref{tab:DWalls} 
and in eq. (\ref{eq:a_diff}) can be used to constrain
mechanisms of SUSY breaking and communication to the visible sector.

The sequence of cosmological events that take place in the two class of models 
is slightly different as indicated in table \ref{tab:pattern}. In both MSLRM and MSLR\rlap/P 
domain walls form when one of the two $\Omega$ fields spontaneously gets a vev 
at a scale $M_R \sim 10^6$ GeV. The domains are distinguished by which of the
two fields acquires a vev in them.
The formation of domain walls results in their energy density dominating 
the energy density of the Universe, in turn causing
a secondary inflation. This dilutes the gravitino and 
moduli fields. As the universe cools the triplet Higgs fields get a vev at a 
scale $M_{B-L} \sim 10^4$ GeV. The parity breakdown mechamism proposed in 
MSLR\rlap/P, causes an asymmetry between $L$ and $R$ sectors  at this
stage. As such DW are de-stabilized and secondary inflation ends.  The motion
of the walls with a preferred direction of motion makes it possible for a 
mechanism for  leptogenesis proposed earlier \cite{cynr02} to be operative 
in MSLR\rlap/P at this energy scale. In MSLRM however, parity is still 
unbroken. At a scale $M_S \sim 10^3$ GeV, SUSY breaking is mediated from 
the hidden sector to the visible sector in both the models. 
The soft terms which come into play at this stage break the parity in 
MSLRM explicitly. Thus, only at this stage do DW become unstable in MSLRM, 
thereby ending secondary inflation and commencing leptogenesis.
The walls finally disappear in MSLR\rlap/P at a temperature of 
$T_D \sim 10 - 10^3$ GeV unlike MSLRM where the same thing happens at a 
temperature range of $T_D \sim 10-10^2$ GeV. While the motion of the walls
produces lepton asymmetry, the walls decay due to collisions dumping entropy 
into the medium and reheating the Universe. For MSLR\rlap/P the reheat 
temperature from secondary inflation $(T^s_R)$ can be estimated to range from 
$10^3 - 10^4$ GeV, whereas for MSLRM,  it can be estimated to be $\lesssim 10^3$ GeV. 
Standard cosmology follows from  this epoch onwards. The estimates of reheating
are based on the energy stored in the walls and accord with the requirement
that sphaleronic processes be effective in converting the lepton asymmetry 
generated by DW mechanism into baryon asymmetry.  The values of reheating 
temperature also keep open the possibly of other TeV scale mechanisms for 
baryogenesis in these models.

In summary we find both MSLRM and MSLR{\rlap/P} having salient cosmological
features which can cure the problems of TeV scale supersymmetric theories.
Both models bear further investigation due the constraints implied
on the mechanism of supersymmetry breaking. The issue of obtaining
non-thermal leptogenesis in this class of models is also an open question.

\section{Acknowledgments}
\label{sec:acknowledge} 
This work is supported by a grant from the Department of Science and
Technology, India. Some of it was completed in the course of a visit
of UAY hosted by Abdus Salam ICTP, Trieste. The work of AS is supported by 
Council of Scientific and Industrial Research, India.

\appendix
\section{$F$-flatness and $D$-flatness conditions}
\label{sec:appendix} 

The $F$-flatness conditions for the MSLR\rlap/P are:
\begin{eqnarray}
F_{\bar\Delta} &=& m_\Delta \Delta 
	+ a (\Delta \Omega - \frac{1}{2} {\rm Tr}\,\Delta\Omega) =0 
\nonumber  \\
F_{\bar{\Delta}_c} &=& m_\Delta  \Delta_c  
	- a (\Delta_c \Omega_c - \frac{1}{2} {\rm Tr}\,\Delta_c\Omega_c)=0
\nonumber  \\ 
F_\Delta &=& m_\Delta  \bar\Delta +i {\bf f} L L^T\tau_2 
	+ a (\Omega\bar\Delta- \frac{1}{2} {\rm Tr}\,\Omega\bar\Delta)=0
\nonumber  \\
F_{\Delta_c} &=& m_\Delta \bar\Delta_c +i {\bf f}
	L_c L_c^T \tau_2
	- a (\Omega_c{\bar\Delta_c}-\frac{1}{2}{\rm Tr}\,
	\Omega_c\bar\Delta_c) =0
\nonumber  \\
F_\Omega &=& m_\Omega \Omega
	+a (\bar\Delta\Delta  - \frac{1}{2} {\rm Tr}\,\bar\Delta\Delta) = 0 
\nonumber \\    
F_{\Omega_c} &=& m_{\Omega} \Omega_c
       	- a (\bar\Delta_c\Delta_c  - \frac{1}{2} {\rm Tr}\,\bar\Delta_c
 \Delta_c) = 0
\nonumber \\
F_{L} &=& 2 i {\bf f} \tau_2 \Delta L = 0
\nonumber \\
F_{L_c} &=& 2 i {\bf f^*} \tau_2 \Delta_c L_c = 0
\label{f-flat}
\end{eqnarray}

The $D$-flatness conditions are given by
\begin{eqnarray}
D_{R i} &= & 2 {\rm Tr}\,\Delta_c^\dagger\tau_i\Delta_c + 2 
{\rm Tr}\,\bar\Delta_c^\dagger\tau_i\bar \Delta_c 
+ 2 {\rm Tr}\,\Omega_c^\dagger\tau_i\Omega_c
+ L_c^\dagger \tau_i L_c = 0 
\nonumber \\
D_{L i} &=&  2 {\rm Tr}\,\Delta^{ \dagger}\tau_i\Delta 
+ 2 {\rm Tr}\,\bar\Delta^{ \dagger}\tau_i\bar \Delta 
+  2 {\rm Tr}\,\Omega^{ \dagger}\tau_i\Omega + L^{\dagger} \tau_i L  = 0 
\nonumber  \\
D_{B-L} &= & -L^\dagger L  + 2 {\rm Tr}\,(\Delta^{ \dagger}\Delta 
- \bar\Delta^{ \dagger}\bar \Delta)
+ L_c^\dagger L_c   - 2 {\rm Tr}\,(\Delta_c^\dagger\Delta_c 
- \bar\Delta_c^\dagger\bar \Delta_c )=0
\label{d-flat}
\end{eqnarray}

\end{document}